\documentclass[twocolumn,british,superscriptaddress, showpacs]{revtex4}
\usepackage[T1]{fontenc}
\usepackage[latin9]{inputenc}
\usepackage{fancyhdr}
\usepackage{graphicx}
\usepackage{esint}

\providecommand{\tabularnewline}{\\}

\usepackage{babel}

\begin{document}

\title{Investigation of the Exclusive $^{3}$H\emph{e(e,e$'$pn)p} Reaction}

\author{D.G. Middleton}

\email{duncan@pit.physik.uni-tuebingen.de}

\affiliation{Kepler Centre for Astro and Particle Physics, Physikalisches Institut,
Universität Tübingen, D-72076 Tübingen, Germany}

\author{J.R.M. Annand}

\affiliation{Department of Physics and Astronomy, University of Glasgow, Glasgow
G12 8QQ, Scotland}

\author{M. Ases Antelo}

\author{C. Ayerbe}

\affiliation{Institut für Kernphysik, Johannes Gutenberg-Universität Mainz, D-55099
Mainz, Germany}

\author{P. Barneo}

\affiliation{Nikhef, P.O. Box 41882, 1009 DB Amsterdam, The Netherlands}

\author{D. Baumann}

\author{J. Bermuth}

\author{J. Bernauer}

\affiliation{Institut für Kernphysik, Johannes Gutenberg-Universität Mainz, D-55099
Mainz, Germany}

\author{H.P. Blok}

\affiliation{Nikhef, P.O. Box 41882, 1009 DB Amsterdam, The Netherlands}

\affiliation{Dept. of Physics, VU-university, Amsterdam, The Netherlands}

\author{D. Bosnar}

\affiliation{Department of Physics, University of Zagreb, Croatia}

\author{R. Böhm}

\author{M. Ding}

\author{M.O. Distler}

\author{J. Friedrich}

\author{J. García Llongo}

\affiliation{Institut für Kernphysik, Johannes Gutenberg-Universität Mainz, D-55099
Mainz, Germany}

\author{D.I. Glazier}

\affiliation{Department of Physics and Astronomy, University of Glasgow, Glasgow
G12 8QQ, Scotland}

\author{J. Golak}

\affiliation{M. Smoluchowski Institute of Physics, Jagiellonian University, PL-30059
Kraków, Poland}

\author{W. Glöckle}

\affiliation{Institut für Theoretische Physik II, Ruhr-Universität Bochum, D-44780
Bochum, Germany}

\author{P. Grabmayr}

\author{T. Hehl}

\author{J. Heim}

\affiliation{Kepler Centre for Astro and Particle Physics, Physikalisches Institut,
Universität Tübingen, D-72076 Tübingen, Germany}

\author{W.H.A. Hesselink}

\affiliation{Nikhef, P.O. Box 41882, 1009 DB Amsterdam, The Netherlands}

\affiliation{Dept. of Physics, VU-university, Amsterdam, The Netherlands}

\author{E. Jans}

\affiliation{Nikhef, P.O. Box 41882, 1009 DB Amsterdam, The Netherlands}

\author{H. Kamada}

\affiliation{Department of Physics, Faculty of Engineering, Kyushu Institute of
Technology, Kitakyushu 804-8550, Japan}

\author{G. Jover Mañas}

\author{M. Kohl}

\affiliation{Institut für Kernphysik, Johannes Gutenberg-Universität Mainz, D-55099
Mainz, Germany}

\author{L. Lapikás}

\affiliation{Nikhef, P.O. Box 41882, 1009 DB Amsterdam, The Netherlands}

\author{I.J.D. MacGregor}

\affiliation{Department of Physics and Astronomy, University of Glasgow, Glasgow
G12 8QQ, Scotland}

\author{I. Martin}

\affiliation{Kepler Centre for Astro and Particle Physics, Physikalisches Institut,
Universität Tübingen, D-72076 Tübingen, Germany}

\author{J.C. McGeorge}

\affiliation{Department of Physics and Astronomy, University of Glasgow, Glasgow
G12 8QQ, Scotland}

\author{H. Merkel}

\author{P. Merle}

\affiliation{Institut für Kernphysik, Johannes Gutenberg-Universität Mainz, D-55099
Mainz, Germany}

\author{K. Monstad}

\affiliation{Department of Physics and Astronomy, University of Glasgow, Glasgow
G12 8QQ, Scotland}

\author{F. Moschini}

\affiliation{Kepler Centre for Astro and Particle Physics, Physikalisches Institut,
Universität Tübingen, D-72076 Tübingen, Germany}

\author{U. Müller}

\affiliation{Institut für Kernphysik, Johannes Gutenberg-Universität Mainz, D-55099
Mainz, Germany}

\author{A. Nogga}

\affiliation{Institute for Advance Simulation, Institut für Kernphysik, and Jülich
Center for Hadron Physics, Forschungszentrum Jülich, D-52425 Jülich,
Germany}

\author{R. P\'{e}rez Benito}

\author{Th. Pospischil}

\affiliation{Institut für Kernphysik, Johannes Gutenberg-Universität Mainz, D-55099
Mainz, Germany}

\author{M. Potokar}

\affiliation{Institute Jožef Stefan, University of Ljubljana, Ljubljana, Slovenia}

\author{G. Rosner}

\affiliation{Department of Physics and Astronomy, University of Glasgow, Glasgow
G12 8QQ, Scotland}

\author{M. Seimetz}

\affiliation{Institut für Kernphysik, Johannes Gutenberg-Universität Mainz, D-55099
Mainz, Germany}

\author{R. Skibi\'{n}ski}

\affiliation{M. Smoluchowski Institute of Physics, Jagiellonian University, PL-30059
Kraków, Poland}

\author{H.~de~Vries}

\affiliation{Nikhef, P.O. Box 41882, 1009 DB Amsterdam, The Netherlands}

\author{Th. Walcher}

\affiliation{Institut für Kernphysik, Johannes Gutenberg-Universität Mainz, D-55099
Mainz, Germany}

\author{D.P. Watts}

\affiliation{Department of Physics and Astronomy, University of Glasgow, Glasgow
G12 8QQ, Scotland}

\author{M. Weinriefer}

\author{M. Weiss}

\affiliation{Institut für Kernphysik, Johannes Gutenberg-Universität Mainz, D-55099
Mainz, Germany}

\author{H. Wita\l{}a}

\affiliation{M. Smoluchowski Institute of Physics, Jagiellonian University, PL-30059
Kraków, Poland}

\author{B. Zihlmann}

\affiliation{Nikhef, P.O. Box 41882, 1009 DB Amsterdam, The Netherlands}

\affiliation{Dept. of Physics, VU-university, Amsterdam, The Netherlands}

\date{\today}
\begin{abstract}
Cross sections for the $^{3}$He\emph{(e,e$'$pn)p} reaction were
measured for the first time at energy transfers of 220 and 270 MeV
for several momentum transfers ranging from 300 to 450 MeV/\emph{c}.
Cross sections are presented as a function of the momentum of the
recoil proton and the momentum transfer. Continuum Faddeev calculations
using the Argonne $V18$ and Bonn-B nucleon-nucleon potentials overestimate
the measured cross sections by a factor 5 at low recoil proton momentum
with the discrepancy becoming much smaller at higher recoil momentum.
\end{abstract}


\maketitle
The understanding of nucleon-nucleon (\emph{NN}) interactions within
the nucleus is of great importance for modern nuclear physics. These
\emph{NN} interactions, which are characterised at short inter-nucleon
separations by a strong scalar repulsive component and at intermediate
to large separations by an attractive part, caused mainly by the strong
tensor component of the meson-exchange contribution, induce correlations
between the nucleons. The use of electron-induced exclusive two-nucleon
knockout reactions of the type A\emph{(e,e$'$pN)}A-2 is a very direct
method for the study of this correlated behaviour within the nucleus.
Because the scalar and tensor interactions act differently in isospin
T~=~0 and T~=~1 states proton-proton and proton-neutron knockout
reactions probe predominantly the short-range and tensor component,
respectively \citep{Schiavilla_PRL_tensor,Wiringa_PRC_78_NN}.

The use of $^{3}$He as a laboratory for the study of \emph{NN} correlations
via \emph{(e,e$'$pN)N} reactions has advantages, both experimentally
and theoretically, over other nuclei. From an experimental point of
view the final state is a single nucleon in its ground state so detector
resolution is not critical and reconstruction of the final state is
straightforward. Furthermore theoretical models exist \citep{eepn_th_meij,Golak_theory_report,3He_deltuva}
that allow the break-up cross section to be calculated exactly, with
interactions between all three nucleons being completely taken into
account. Such models calculate both the $^{3}$He ground state and
the three-nucleon continuum wave functions using realistic \emph{NN}
potentials which include a phenomenological description of the short
range part of the \emph{NN }interaction \citep{few_nucl_syst_rev,Nogga_PRC.65.054003}.

At electron energies of several hundred MeV the electron-induced two-nucleon
knockout cross section is driven by several processes. The coupling
of the virtual photon to one nucleon of a correlated pair via one-body
hadronic currents can lead to the ejection of both nucleons from the
nucleus. Interaction of the virtual photon with two-body hadronic
currents, such as meson exchange currents (MEC) or isobar currents
(IC), also contributes to the cross section. In addition to the above
processes there can also be interactions between all particles in
the final state (FSI), the strength of which depends strongly on the
relative nucleon-nucleon energies. The relative importance of these
different processes depends on the kinematics and type of reaction.
In order to disentangle these different contributing processes it
is important to measure both the \emph{(e,e$'$pp)} and \emph{(e,e$'$pn)}
cross sections as a function of several kinematic variables. The $^{3}$He\emph{(e,e$'$pp)n}
reaction was studied previously in \citep{3he_eepp_let,3He_eepp,3he_eepp_clas}.
Here we present the results of a measurement of the $^{3}$He\emph{(e,e$'$pn)p}
reaction made in the so-called {}``dip-region'' between quasi-elastic
and $\Delta$-excitation peaks in the inclusive electron-scattering
spectrum.

The measurements were performed at the electron scattering facility
of the 100\% duty factor Mainz Microtron MAMI \citep{MAMI_B_design,MAMI_B_Walcher}.
The 855 MeV electron beam, used with currents between 2 and 4 $\mu$A,
was incident on a $^{3}$He high-pressure cryogenic gas target, operated
at 1.9 MPa and 15 K. At 4 $\mu$A beam current this corresponds to
a luminosity of $2\times10{}^{36}$~cm$^{-2}$~s$^{-1}$. The scattered
electrons were detected in Spectrometer B \citep{SpecB}, a magnetic
spectrometer with a solid angle of $\Delta\Omega$ = 5.6 msr and momentum
acceptance of $\Delta p/p$ = 15\%. The ejected protons were detected
using the scintillator detector HADRON3 (H3) \citep{HADRON3} from
Nikhef, a large solid angle ($\Delta\Omega$ = 230 msr) hodoscope
consisting of 128 bars of plastic scintillator divided into eight
layers: two hodoscope layers at the front with six energy-determining
layers behind. The proton energy acceptance of H3 is 50 - 250 MeV.
For detection of the ejected neutrons the Glasgow-Tübingen time-of-flight
(TOF) detector system \citep{TOF} was used. The TOF detector array
consisted of 96 bars of plastic scintillator arranged in three stands
of 32 bars. Each stand consisted of 4 layers of 8 detectors, 3 layers
of 5 cm thick TOF bars positioned behind a layer of 8 overlapping
1 cm thick veto detectors in a configuration similar to that described
in \citep{16O_eepn_exp}. The TOF array covered a solid angle of $\Delta\Omega$
$\approx$ 240 msr. 

Kinematics were chosen so that the reaction could be studied over
a range of energy and momentum transfers similar to that covered previously
in a measurement of the $^{3}$He\emph{(e,e'pp)n} reaction \citep{3He_eepp}.
The settings used for each of the different kinematics studied are
summarised in table \ref{tab:Kinematics}. Data were taken at momentum
transfers of \emph{q}~=~300 and 375 MeV/\emph{c} for an energy transfer
of $\omega$~=~220~MeV and at momentum transfers of \emph{q}~=~330,
375 and 450 MeV/\emph{c} for an energy transfer of $\omega$~=~270~MeV.
The proton detector, H3, was positioned in the direction of
the momentum transfer $\vec{q}$ at forward angles with respect to
the beam on the opposite side of the beam line to Spectrometer B while
the TOF neutron detectors were positioned at $180^{\circ}$ to the direction
of $\vec{q}$ at backward angles with respect to the beam on the same
side as Spectrometer B. After software cuts the neutron energy threshold
in TOF was 16.7~MeV, while the use of 0.5~cm thick lead shielding
in front of H3 resulted in an effective proton energy threshold of
70~MeV. %
\begin{table*}
\begin{centering}
\begin{tabular}{|c|c|c|c|c|c|c|c|c|}
\hline 
Label & $\omega$ & \emph{q} & $\theta_{B}$ & $\theta_{H3}$ & $\theta_{TOF-1}$ & $\theta_{TOF-2}$ & $\theta_{TOF-3}$ & $\theta_{q}$\tabularnewline
 & {[}MeV{]} & {[}MeV/\emph{c}{]} & {[}deg{]} & {[}deg{]} & {[}deg{]} & {[}deg{]} & {[}deg{]} & {[}deg{]}\tabularnewline
\hline 
A1 & 220 & 375 & 23.8 & -53.3 & 107.7 & 125.5 & 140.5 & -43.1\tabularnewline
\hline 
A2 & 270 & 375 & 20.9 & -44.0 & 107.7 & 125.5 & 140.5 & -34.3\tabularnewline
\hline 
B2 & 270 & 450 & 29.3 & -44.0 & 107.7 & 125.5 & 140.5 & -39.8\tabularnewline
\hline 
Y1 & 220 & 300 & 15.4 & -45.0 & 107.7 & 125.5 & 140.5 & -35.5\tabularnewline
\hline 
Z2 & 270 & 330 & 15.9 & -45.0 & 107.7 & 125.5 & 140.5 & -28.1\tabularnewline
\hline
\end{tabular}
\par\end{centering}

\caption{\label{tab:Kinematics}Kinematic settings in which the data were taken.
The detector angles given are the central angle of each detector in
the lab frame. }

\end{table*}

With the detection of two ejected nucleons from the initial 3N-system,
the kinematics of the reaction are completely determined. The missing
momentum of the reaction is defined as 

\begin{equation}
\vec{p}_{m}=\vec{q}-\vec{p}_{p'}-\vec{p}_{n'}\label{eq:pm}\end{equation}
and is equal to the momentum of the undetected proton, $\vec{p}_{r}$;
here $\vec{p}_{p'}$ and $\vec{p}_{n'}$ are the momenta of the detected
proton and neutron, respectively. Using $\vec{p}_{m}$, the missing-energy

\begin{equation}
E_{m}=\omega-T_{p'}-T_{n'}-T_{r'}\label{eq:em}\end{equation}
can be determined where $T_{p'}$, $T_{n'}$ and $T_{r'}$ are the
kinetic energies of the proton, neutron and undetected proton, respectively. 

\begin{figure}
\begin{centering}
\includegraphics[width=1\columnwidth]{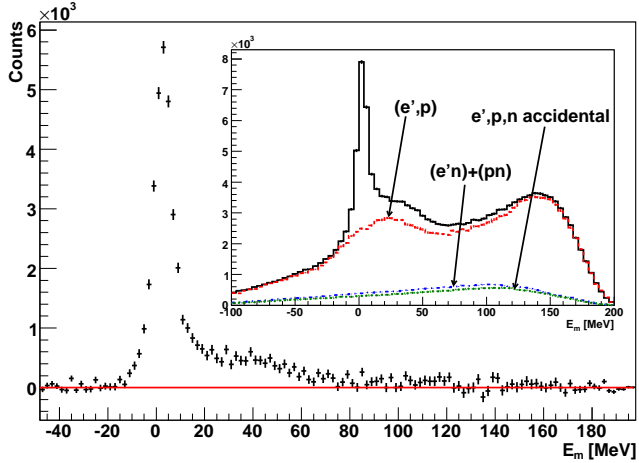}
\par\end{centering}

\caption{\label{Z2_Em} (colour online). The missing-energy ($E_{m}$) distribution
for the $^{3}$He\emph{(e,e$'$pn)} reaction for the Z2 kinematic
setting. The inset shows the $E_{m}$ distribution before subtraction
of accidental coincidences. The black curve results from all \emph{(e$'$pn)}
events in a coincidence time region where real triple-coincidence
events would be found. The dashed red curve shows the contribution
from events with an accidental neutron, the dot-dashed blue curve
shows the contribution from events with an accidental proton or electron
and the dotted green line shows the contribution from events where
all three particles are in accidental coincidence.}

\end{figure}

The missing-energy spectrum for the Z2 kinematic setting is shown
in Fig. \ref{Z2_Em}. The inset of Fig. \ref{Z2_Em} shows the missing-energy
spectrum before the subtraction of accidental coincidences as well
as the spectra for the different types of accidental coincidences.
The \emph{(e$'$p)} and \emph{(e$'$n)+(pn)} types of accidental coincidences
(red dashed and blue dotted curves) were subtracted from the total
coincidence yield (black curve). As this subtracts the ``\emph{e$'$,p,n}
accidental'' coincidences twice (green dot-dashed curve) these were
added to the result to obtain the number of real coincident events.
More details of the accidental subtraction procedure can be found
in reference \citep{16O_eepn_exp}. The largest contribution to the
background is from events consisting of a real coincident \emph{(e$'$p)}
pair and an accidental neutron. The missing-energy spectrum corrected
for accidental coincidences has a single peak which corresponds to
the three-body break-up of $^{3}$He. The peak has a FWHM of 7.0 MeV,
as expected from the known detector resolutions in energy and angle,
and a mean value of 6.2 MeV, close to the expected value of 7.72 MeV.
The tail at higher missing energies is due to radiative processes.%

The 8-fold differential cross section for the $^{3}$He\emph{(e,e$'$pn)}
reaction can be written as a 5-fold differential cross section, $d^{5}\sigma/dS\, d\Omega_{p}d\Omega_{n}$,
containing the nuclear structure information, multiplied by a virtual-photon
flux factor $\Gamma_{v}$ \citep{virtual_photon_flux}. The cross
section is differential in\emph{ S}, where $dS$ is the arc-length
in the $T_{p'}-T_{n'}$ plane along the curve\emph{ S} that describes
the relation between $T_{p'}$ and $T_{n'}$ for a given proton-neutron
angular configuration as given by Eq. \ref{eq:em} \citep{3He_ppn_calcs,dS_definition}.
It depends on seven independent kinematic variables, but because of
the limited statistics of the data, it will be presented as function
of one of them only, integrating over the others within the acceptance
of the detectors. The average 5-fold experimental cross section was
calculated as:

\begin{equation}
\frac{d^{5}\sigma}{dS\, d\Omega_{p}d\Omega_{n}}(x)=\frac{1}{\mathcal{\int L}dt}\frac{N(x,\Delta x)}{\mathcal{V}(x,\Delta x)}\label{eq:d5_sigma}\end{equation}
where $\Delta x$ represents a range (bin) in the variable $x$ as
a function of which the cross section is presented, $\mathcal{\int L}dt$
is the integrated luminosity, $N(x,\Delta x)$ is the measured number
of \emph{(e,e$'$pn)} events in bin $\Delta x$, corrected for accidental
coincidences, integrated over the missing-energy range from below
the peak up to 36 MeV, and $\mathcal{V}(x,\Delta x)$ is the corresponding
(weighted) detection volume. The latter was determined by a Monte
Carlo method using 10$^{8}$ events, generated within the energy/momentum
acceptances of the three detectors involved, taking into account energy
conservation for the reaction. This yielded an eight-dimensional phase-space
or detection volume $\mathcal{V}$. Radiative corrections were applied
to the phase space which were calculated using the formalism of Mo
and Tsai \citep{radiative_corrections}. The factor $\Gamma_{v}$
was also included as a weight, as were the efficiencies of the H3
and TOF detectors. The former ranged from $90\%$ to $80\%$ depending
on the proton energy. The neutron detection efficiency was about $2.7\%$
on average for a 5 cm thick TOF bar at the signal pulse-height threshold
used. This energy dependent efficiency was computed with a model based
on the Stanton code\emph{ }\citep{stanton,Cecil}. Finally the phase
space was integrated over the missing-energy to yield $\mathcal{V}(x,\Delta x)$
using the same limits as in the determination of $N(x,\Delta x)$,
and over all variables except \emph{x}, applying the same cuts in
$T_{p'}$ and $T_{n'}$ as for the experimental data. The statistical
error associated with the generated phase-space is $\leq0.5\%$. Corrections
for dead time in the electronics were included in the determination
of the integrated luminosity $\mathcal{\int L}dt$.

In the figures only the statistical errors are shown. The overall
systematic error is about 13\% with the largest contribution coming
from the uncertainty ($\approx12\%$) in the neutron detection efficiency.
The uncertainty in the correction for hadronic interactions and multiple
scattering in H3 is about 4\%. Other contributions to the systematic
error such as those from luminosity calibration, dead time corrections
and target-thickness determination from elastic scattering, are negligible
compared to those of the detection efficiency corrections.

The measured cross sections are compared to the results of non-relativistic
continuum Faddeev calculations \citep{Golak_theory_report} using
the Argonne $V18$ and Bonn-B nucleon-nucleon potentials. The calculations
contain mechanisms for photon absorption on one, two or all three
of the nucleons in the target but are limited in their treatment of
exchange and isobar currents. The calculations are only strictly applicable
for photon energies below pion production threshold. Rescattering
processes up to all orders in the continuum are included ensuring
that all FSI effects are fully taken into account. Two types of calculations
were made. The first employed only a one-body hadronic current operator,
while the second also included a two-body current operator for $\pi$
and $\rho$ mesons to account for the in-flight and seagull terms
using the formalism of Schiavilla \emph{et al.} \citep{MEC_currents_for_calcs}
which is based on earlier work of Riska \citep{riska_mec_1,riska_mec_2}. 

For comparison to the measured data the calculation of the theoretical
cross sections was done in two steps. In the first part the Faddeev
equations were solved for a single, central \emph{($\omega$,q)} point
per kinematic setting. All \emph{NN}-force components were included
up to a two-body angular momentum of $j=3$ and in the final state
all partial waves were included up to three-body angular momentum
$J=\frac{15}{2}$. Then, in the second step, the cross section was
calculated for many specific three-nucleon final states determined
by $\vec{p}_{p'}$ and $\vec{p}_{n'}$ for a given \emph{($\omega$,q)}
point. The final states were randomly generated over the full acceptance
of the H3 and TOF detectors using a generator similar to that used
for the phase space. Approximately $10^{7}$ events were generated
for each \emph{($\omega$,q)} point to ensure full coverage and to
reduce statistical fluctuations. Then the average cross section as
a function of one variable integrated over all the others was calculated.
This method of determining the theoretical cross section for a given
experimental set-up is more accurate than that used previously for
comparison to the measurement of the $^{3}$He(\emph{e,e$'$pp}) reaction
\citep{3He_eepp} where the cross sections were calculated for a grid
of points covering the energy and angular acceptances of the two proton
detectors.

The effect of using only the central\emph{ ($\omega$,q)} point was
investigated for one kinematic setting. The average difference in
the calculated theoretical cross section between using just the central
value or using a range covering the whole acceptance of spectrometer
B was found to be about 10\%.

\begin{figure}
\begin{centering}
\includegraphics[width=1\columnwidth]{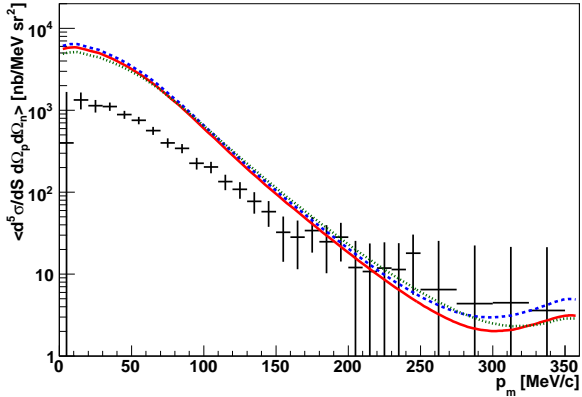}
\par\end{centering}

\caption{\label{A1_pm} (colour online). The cross section for the $^{3}$He\emph{(e,e$'$pn)}
reaction shown as a function of the missing momentum for the A1 kinematic
setting. The solid red (dotted green) curve shows the theoretical
cross section calculated using only a one-body hadronic current operator
and the Argonne $V18$ (Bonn-B) \emph{NN} potential. The dashed blue
line results from the Argonne potential when MECs are also included.}

\end{figure}

Fig. \ref{A1_pm} shows the $^{3}$He\emph{(e,e$'$pn)} cross section
as a function of the  missing momentum for the A1 kinematic setting.
The solid red line shows the theoretical cross section calculated
using the Argonne $V18$ \emph{NN} potential and a one-body hadronic
current operator, while the dashed blue line shows the results when
MECs are also included. The dotted green line in Fig. \ref{A1_pm}
shows the result of the one-body calculation using the Bonn-B \emph{NN}
potential. %

In Fig. \ref{A1_pm} the overall shape of the measured and theoretical
cross sections is similar in that they decrease roughly exponentially
with increasing $p_{m}$. The inclusion of the MECs increases the
calculated cross section by only about 10\% up to $p_{m}$ = 200 MeV/\emph{c}
but the effect increases to about 60\% at 350 MeV/\emph{c}. The use
of the two different potentials makes little difference to the results
of the calculations which overpredict the experimental data by about
a factor of 5 for $p_{m}$ $\leq$ 80 MeV/\emph{c}. The discrepancy
decreases with increasing $p_{m}$ until rough agreement within the
large experimental error bars is reached at $p_{m}$ $\approx$ 200
MeV/\emph{c}. In general the calculations indicate that for the kinematics
shown here the $^{3}$He\emph{(e,e$'$pn)} cross section is dominated
by the one-body hadronic current term with MECs only making a minor contribution.

The overprediction of the cross section by the theoretical calculations
is in contrast to the $^{3}$He\emph{(e,e$'$pp)} reaction \citep{3He_eepp}
where calculations using the Bonn-B potential are slightly below the
data at low $p_{m}$ and a factor of 5 lower at $p_{m}=200$ MeV/\emph{c}.
There is not only a clear difference in the ratio between the \emph{pp}
and \emph{pn}-knockout data and their respective one-body current
prediction, but also the measured cross sections do not fall as quickly
as predicted with increasing $p_{m}$. 

This different behaviour between \emph{pp} and \emph{pn} knockout is intriguing. The large discrepancy at low $p_{m}$ in the case of \emph{pn} knockout, where the virtual photon supposedly couples mainly to one of the nucleons of a \emph{pn} pair, would suggest that the \emph{pn} correlations in the probed regime are not well predicted by the theory. Another source of disagreement could be the incomplete treatment of MECs and the omission of ICs, which play a much larger role in \emph{pn} than in \emph{pp} knockout. Finally, as stated above, the theory is only applicable for photon energies below pion production threshold.

\begin{figure}
\begin{centering}
\includegraphics[width=1\columnwidth]{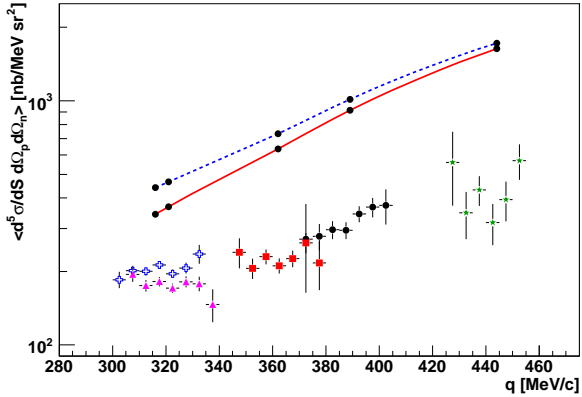}
\par\end{centering}

\caption{\label{q_xsec} (colour online). The cross section for the $^{3}$He\emph{(e,e$'$pn)}
reaction shown as a function of \emph{q} for 50 $\leq p_{m}\leq$100
MeV/\emph{c} and 235 $\leq\omega\leq$ 265 MeV. Results from different
kinematic settings are shown by the different types of markers for
the data points: A1:black circles; A2: red squares; B2: green stars;
Y1: magenta triangles; Z2: blue crosses. The same line convention
is used for the theoretical curves as in Fig. \ref{A1_pm}. The theoretical
cross section was calculated at the points in the curves indicated
by the black circles.}

\end{figure}

Fig. \ref{q_xsec} shows the $^{3}$He(\emph{e,e$'$pn}) cross section
as a function of \emph{q} for the $\omega$ range 235 $\leq\omega\leq$
265 MeV and the $p_{m}$ range 50 $\leq p_{m}\leq$ 100 MeV/\emph{c}.
The experimental cross section in Fig. \ref{q_xsec} was measured
in five different kinematic settings, shown by the different types
of markers for the data points, and shows a smoothly rising \emph{q-}dependence
which increases by about a factor 2 in strength from \emph{q} = 300
to 450 MeV/\emph{c}. The line convention for the theoretical curves
is the same as for Fig. \ref{A1_pm}. The theoretical cross sections
were calculated for five separate \emph{q}-values which are indicated
by the filled black circles in the curves. The theoretical calculations
over-predict the measured cross sections for the whole \emph{q} range
covered; by a factor 2-3 at \emph{q} = 320 MeV/\emph{c} to a factor
of about 5 at \emph{q} = 450 MeV/\emph{c}. The inclusion of the MECs
increases the calculated cross section by about 30\% at \emph{q} =
320 MeV/\emph{c}. This increase falls with increasing \emph{q} to
about 5\% at \emph{q} = 450 MeV/\emph{c}. %

The \emph{q}-dependence of the $^{3}$He\emph{(e,e$'$pp)} data for
a similar $p_{m}$ range is much better described by the calculations
with just a \emph{q}-independent under-prediction of 20\%. Again this
points to an inadequacy in the theoretical calculations for \emph{pn}
knockout, as discussed above.

In conclusion the $^{3}$He\emph{(e,e$'$pn)} reaction was measured
with good statistical accuracy over a range of momentum transfers
and for two energy transfers. Calculations using the Argonne $V18$
or Bonn B potentials over-predict the measured cross sections by a
factor 5 at low $p_{m}$ but are in rough agreement within the large
statistical experimental errors at $p_{m}\geq200$~MeV/\emph{c}.
Inclusion of MECs increases the calculated cross section section by
a small amount (about 10\%) up to $p_{m}=200$ MeV/\emph{c}, increasing
to about 60\% at $p_{m}=350$ MeV/\emph{c}. When the cross section
at low $p_{m}$ is considered as a function of \emph{q} the calculations
over-predict the data by a factor 2 to 5.

Comparison with data measured for the $^{3}$He\emph{(e,e$'$pp)}
reaction, which are much better described by the same theoretical
calculations at low $p_{m}$, suggests that the \emph{pn} correlations
in the probed kinematical regime are not well described and/or reaction
mechanisms that are not included in the calculations (such as isobar
currents and certain meson-exchange currents) play a large role.

\begin{acknowledgments}
The authors would like to thank the staff of the Institut für Kernphysik
in Mainz for providing the facilities for this experiment. This work
was sponsored by the UK Engineering and Physical Sciences Research
Council (EPSRC), the Deutsche Forschungsgemeinschaft (DFG) and the
Foundation for Fundamental Research of Matter (FOM), which is financially
supported by the Netherlands Organisation for Scientific Research
(NWO). It was also partially supported by the 2008-2011 Polish Science
Funds as a research project No. NN202 077435. The numerical calculations
were performed at the NIC, Jülich, Germany.
\end{acknowledgments}

\bibliographystyle{/usr/share/texmf/bibtex/bst/revtex/apsrev}

\end{document}